\documentclass[]{spie}

\usepackage{mathptm}
\usepackage{graphicx}
\usepackage{amsfonts}

\title{Graph-based simulation of quantum computation in the density matrix representation}

\author{George F. Viamontes, Igor L. Markov, John P. Hayes
\skiplinehalf
University of Michigan, Advanced Computer Architecture Laboratory, \\ 1301 Beal Ave., Ann Arbor, MI, USA 48109-2122}

\authorinfo{E-mail contact info: \{gviamont, imarkov, jhayes\}@eecs.umich.edu}

\begin{document}
  \maketitle

  \begin{abstract}

    Quantum-mechanical phenomena are playing an increasing role in
    information processing, as transistor sizes approach the nanometer
    level, and quantum circuits and data encoding methods appear in
    the securest forms of communication.  Simulating such phenomena
    efficiently is exceedingly difficult because of the vast size of
    the quantum state space involved.  A major complication is caused
    by errors (noise) due to unwanted interactions between the quantum
    states and the environment.  Consequently, simulating quantum
    circuits and their associated errors using the density matrix
    representation is potentially significant in many applications,
    but is well beyond the computational abilities of most classical
    simulation techniques in both time and memory resources. The size
    of a density matrix grows exponentially with the number of qubits
    simulated, rendering array-based simulation techniques that
    explicitly store the density matrix intractable. In this work, we
    propose a new technique aimed at efficiently simulating quantum
    circuits that are subject to errors. In particular, we describe
    new graph-based algorithms implemented in the simulator
    QuIDDPro/D. While previously reported graph-based simulators
    operate in terms of the state-vector representation, these new
    algorithms use the density matrix representation. To gauge the
    improvements offered by QuIDDPro/D, we compare its simulation
    performance with an optimized array-based simulator called
    QCSim. Empirical results, generated by both simulators on a set of
    quantum circuit benchmarks involving error correction, reversible
    logic, communication, and quantum search, show that the
    graph-based approach far outperforms the array-based approach.

  \end{abstract}

  \keywords{Quantum circuits, quantum algorithms, simulation, density matrices, quantum errors, graph data structures, decision diagrams, QuIDDs}
  \vspace{-2mm}

  \section{INTRODUCTION}

  Practical information-processing applications that exploit
  quantum-mechanical effects are becoming common. For example, MagiQ
  Technologies markets a quantum communications device that detects
  eavesdropping attempts and prevents them\cite{magiq}. The act of
  eavesdropping can be modeled as both making a quantum measurement
  and corruption by environmental
  noise\cite{Nielsen2000}. Additionally, quantum computational
  algorithms have been discovered to quickly search unstructured
  databases\cite{Grover97} and to factor numbers in polynomial
  time.\cite{Shor97} Implementing quantum algorithms has proved to be
  particularly difficult, however, in part due to errors caused by the
  environment\cite{Kielp2002,Monroe2002,Nielsen2000}. Another related
  application is reversible logic circuits. Since the operations
  performed in quantum computation must be unitary, they are all
  invertible and allow re-derivation of the inputs given the
  outputs.\cite{Nielsen2000} This phenomenon gives rise to a host of
  potential applications in fault-tolerant and low-power
  computation. Since reversible logic, quantum communication, and
  quantum algorithms can be modeled as quantum
  circuits\cite{Nielsen2000}, quantum circuit simulation incorporating
  errors could be of major benefit to these applications. In fact, any
  quantum-mechanical phenomenon with a finite number of states can be
  modeled as a quantum circuit\cite{BT97,Nielsen2000}, which may lead
  to other design applications for quantum circuit simulation in the
  future.

  We present a new technique that facilitates efficient simulation of
  the density matrix representation of quantum circuits. The density
  matrix representation is crucial in capturing interactions between
  quantum states and the environment, such as noise. In addition to
  the standard set of operations required to simulate with the
  state-vector model, including matrix multiplication and the tensor
  product, simulation with the density matrix model requires the outer
  product and the partial trace. The outer product is used in the
  initialization of qubit density matrices, while the partial trace
  allows a simulator to differentiate qubit states coupled to noisy
  environments or other unwanted states. The partial trace is
  invaluable in error modeling since it facilitates descriptions of
  single qubit states that have been affected by noise and other
  phenomena.\cite{Nielsen2000}

  Unfortunately, like the state-vector model, simulation with the
  density matrix is computationally challenging on classical
  computers. The size of any density matrix grows exponentially with
  the number of qubits or quantum states it
  represents.\cite{Nielsen2000} Thus, simulation techniques which
  require explicit storage of the density matrix in a series of arrays
  are inefficient and generally intractable. However, the new
  simulation technique we propose is founded in graph-based algorithms
  which can represent and manipulate density matrices very efficiently
  in many important cases. A key component of our algorithms is the
  {\em Quantum Information Decision Diagram} (QuIDD) data structure,
  which can represent and manipulate a useful class of matrices and
  vectors commonly found in quantum circuit applications using time
  and memory resources that are {\em polynomial} in the number of
  qubits\cite{aspdac,tech_report}. A limitation of our previous QuIDD
  algorithms, and other graph-based techniques\cite{shornuf}, is that
  they simulate the state-vector representation of quantum
  circuits. In this work, we present new algorithms to perform the
  outer product and the partial trace with QuIDDs. These algorithms
  enable QuIDD-based simulation of quantum circuits with the density
  matrix representation.

  We also describe a set of quantum circuit benchmarks that
  incorporate errors, error correction, reversible logic,
  communication, and quantum search. To empirically evaluate the
  improvements offered by our new technique, we use these benchmarks
  to compare QuIDD\-Pro/D with an optimized array-based density matrix
  simulator called QCSim.\cite{qcsim} Performance data from both
  simulators show that our new graph-based algorithms far outperform
  the array-based approach.

  The paper is organized as follows. Section \ref{sec:background}
  provides background on decision diagram data structures and previous
  simulation work. In Section \ref{sec:sims} we present our new
  algorithms along with a description of the QuIDD\-Pro/D
  simulator. Section \ref{sec:exp} describes the quantum circuit
  benchmarks and presents performance results on each benchmark for
  QuIDD\-Pro/D and QCSim. Finally, in Section \ref{sec:conclusions} we
  present our conclusions and ideas for future work.

  \section{BACKGROUND AND PREVIOUS WORK}
  \label{sec:background}
  
  The simulation technique proposed in this work relies on the QuIDD
data structure, which is a type of graph called a decision
diagram. This section presents the basic concepts of decision
diagrams, assuming only a rudimentary knowledge of computational
complexity and graph theory. It then reviews previous research on
simulating quantum circuits.

\subsection{Binary Decision Diagrams}
\label{sec:bdd}

\begin{figure}[!htb]
\vspace{2mm}
  \begin{center}
    \begin{tabular}{c|c|c|c}
      \parbox{3cm}{
    \begin{center}
      {\large $f=x_0 \cdot x_1 + x_1$ }
    \end{center}
      }
      &
      \includegraphics[width=4cm]{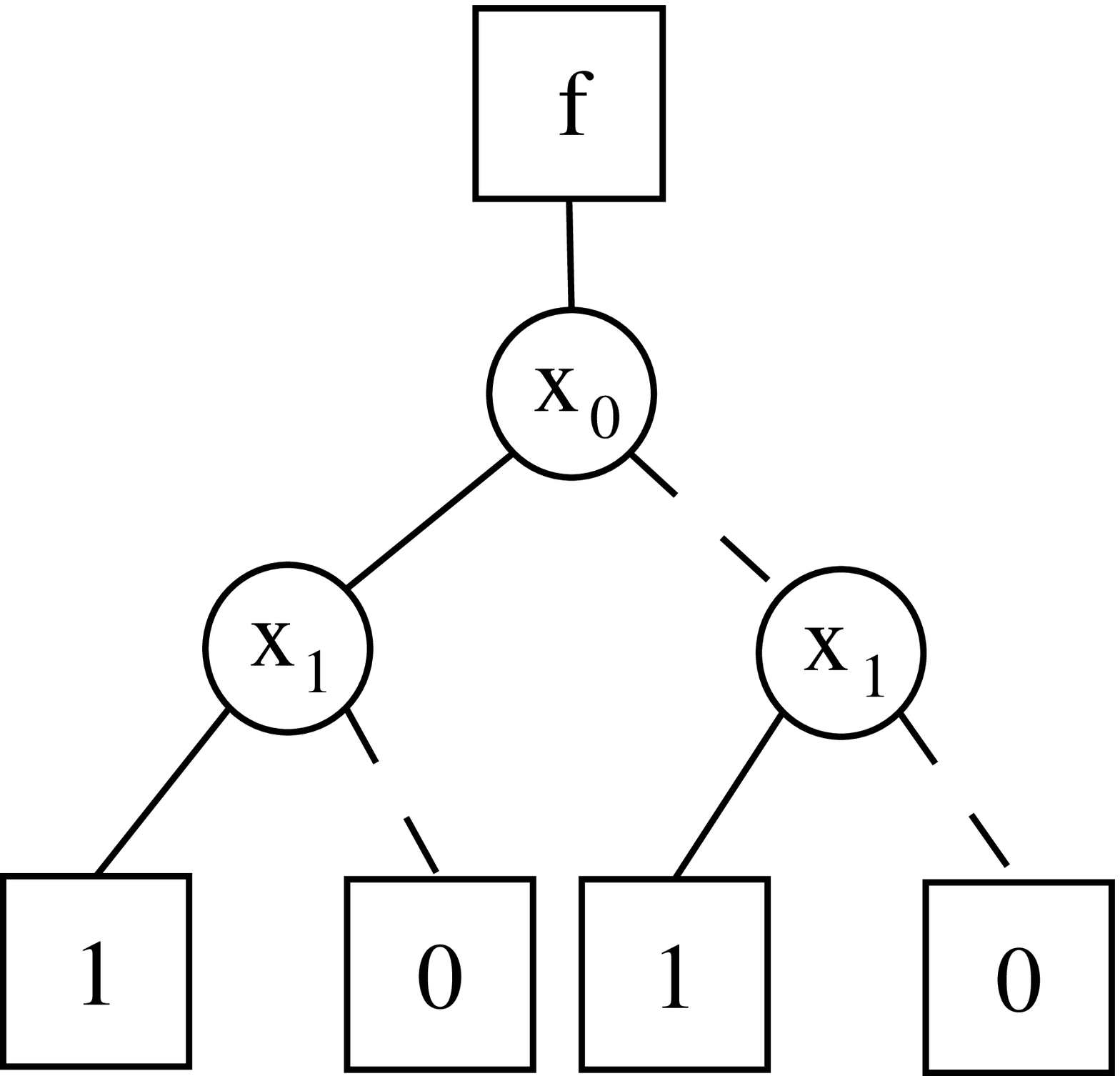}
      &
      \includegraphics[width=2cm]{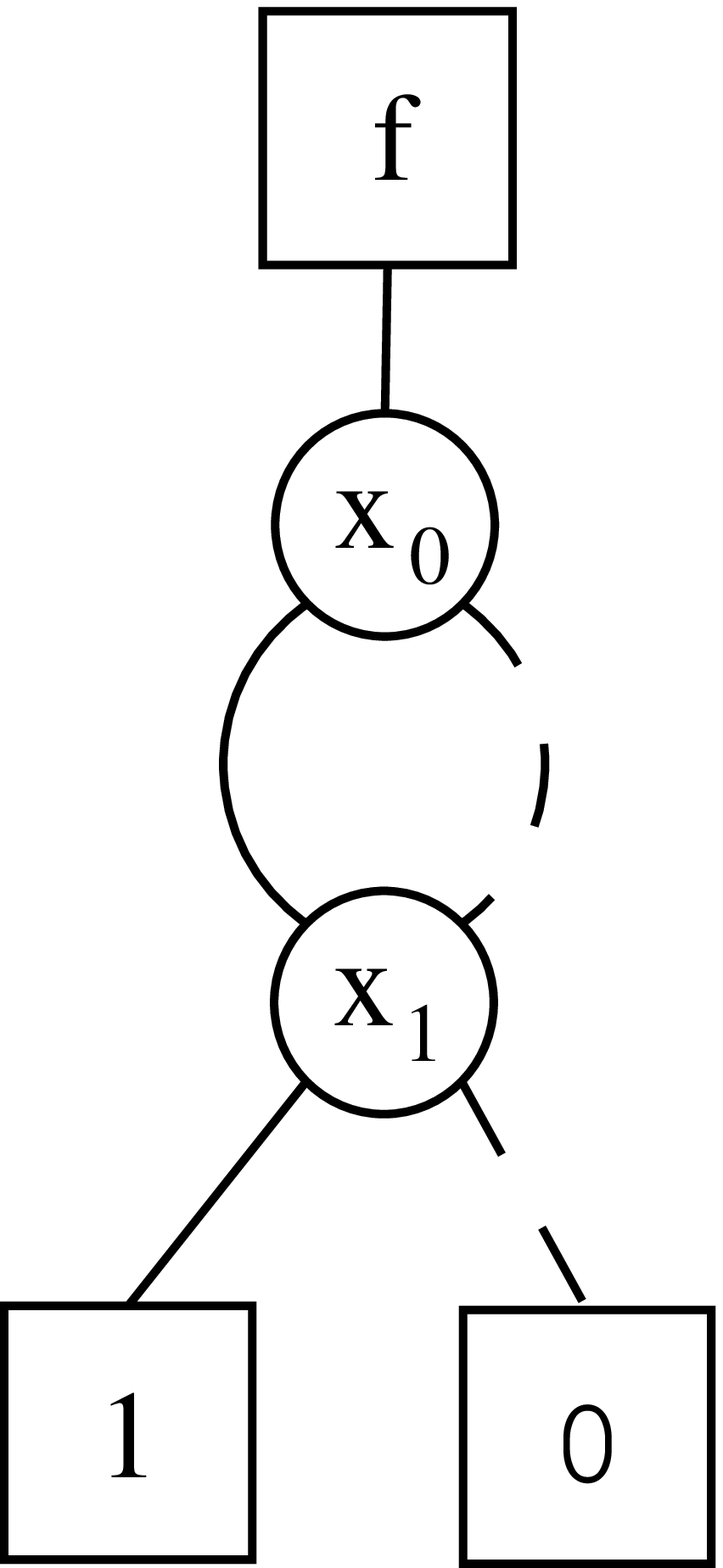}
      &
      \includegraphics[width=2cm]{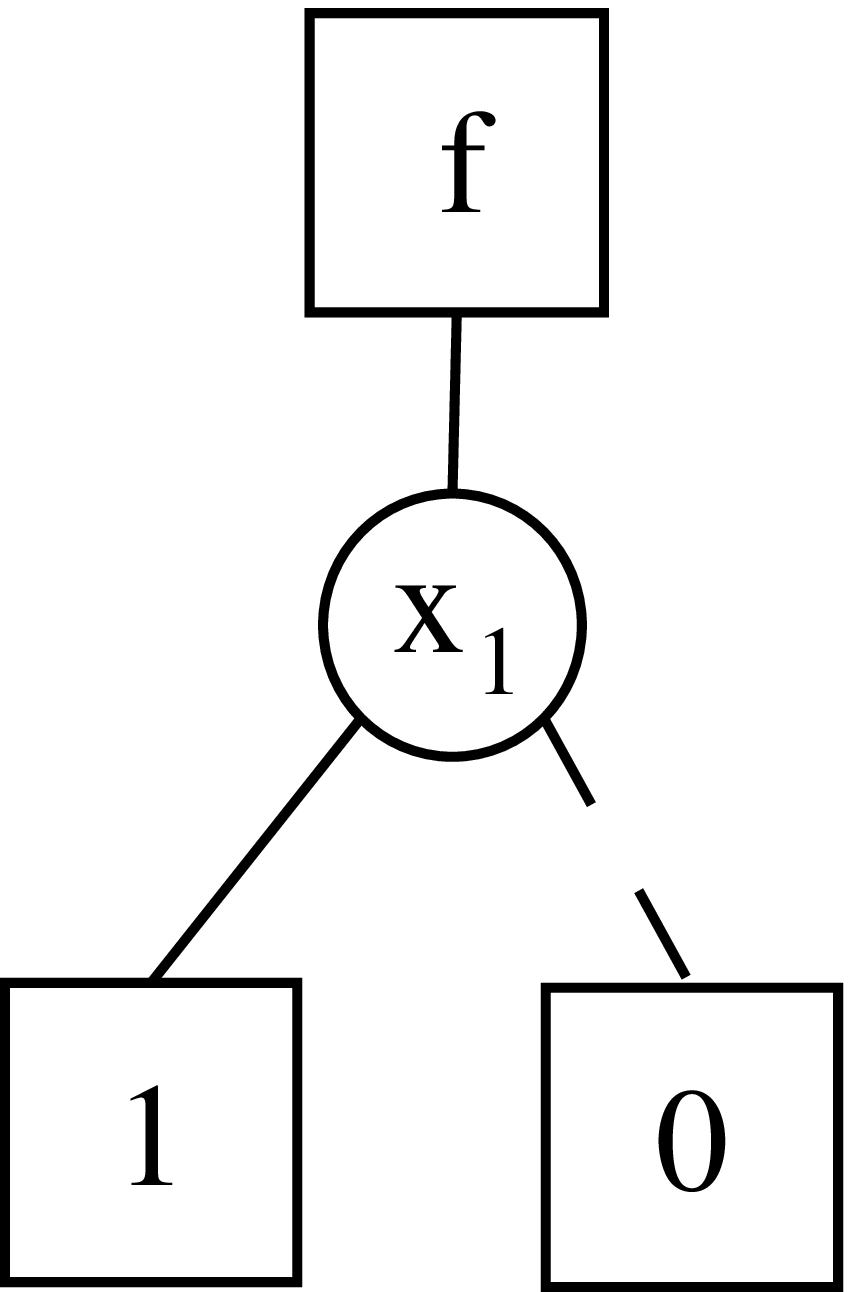}
      \\
      (a) & (b) & (c) & (d)
    \end{tabular}
    \parbox{14cm}{\caption{\label{fig:bdd} (a) A logic function, (b)
    its BDD representation, (c) its BDD representation after applying
    the first reduction rule, and (d) its ROBDD representation.}}
  \end{center}
\vspace{-2mm}
\end{figure}

Many decision diagrams are ultimately based on the binary decision
diagram (BDD). The BDD was developed by Lee in 1959 in the context of
classical logic circuit design.\cite{Lee59} This data structure
represents a Boolean function $f(x_1,x_2,...,x_n)$ by a directed
acyclic graph (DAG) as shown in Fig. \ref{fig:bdd}. By convention, the
top node of a BDD is labeled with the name of the function $f$
represented by the BDD. Each variable $x_i$ of $f$ is associated with
one or more nodes, each of which have two outgoing edges labeled {\em
then} (solid line) and {\em else} (dashed line). The {\em then} edge
of node $x_i$ denotes an assignment of logic $1$ to the $x_i$, while
the {\em else} edge denotes an assignment of logic $0$. These nodes
are called {\em internal} nodes and are labeled by the corresponding
variable $x_i$. The edges of the BDD point downward, implying a
top-down assignment of values to the Boolean variables depicted by the
internal nodes.

At the bottom of a BDD are {\em terminal} nodes containing the logic
values $1$ or $0$. They denote the output value of the function $f$
for a given assignment of its variables. Each path through the BDD
from top to bottom represents a specific assignment of 0-1 values to
the variables $x_1,x_2,...,x_n$ of $f$, and ends with the
corresponding output value $f(x_1,x_2,...,x_n)$.

The memory complexity of the original BDD data structure conceived by
Lee is exponential in the number of variables for a given logic
function. Simulation of many practical logic circuits with this data
structure was therefore impractical. To address this limitation,
Bryant developed the Reduced Ordered BDD (ROBDD) \cite{bryant}, where
all variables are ordered, and assignment of values to variables are
made in that order. A key advantage of the ROBDD is that
variable-ordering facilitates an efficient implementation of reduction
rules that automatically eliminate redundancy from the basic BDD
representation. These rules are summarized as follows:

\begin{enumerate}

  {\item There are no nodes $v$ and $v'$ such that the subgraphs
  rooted at $v$ and $v'$ are isomorphic}

  {\item There are no internal nodes with {\em then} and {\em else}
  edges that both point to the same node}

\end{enumerate}

An example of how the rules distinguish an ROBDD from a BDD is shown
in Fig. \ref{fig:bdd}. The subgraphs rooted at the $x_1$ nodes in
Fig. \ref{fig:bdd}b are isomorphic. By applying the first reduction
rule, the BDD in Fig. \ref{fig:bdd}b becomes the BDD in Fig.
\ref{fig:bdd}c. Note that, in Fig. \ref{fig:bdd}c, the {\em then} and
{\em else} edges of the $x_0$ node now point to the same
node. Applying the second reduction rule eliminates the $x_0$ node,
resulting in the ROBDD in Fig. \ref{fig:bdd}d. Intuitively it makes
sense to eliminate the $x_0$ node since the output of the original
function is determined solely by the value of $x_1$. In many Boolean
functions, this type of redundancy is eliminated with varying success
depending on the order in which variables in the function are
evaluated. Finding the optimal variable ordering is an $NP$-complete
problem, but efficient ordering heuristics have been developed for
specific applications. Moreover, it turns out that many practical
logic functions have ROBDD representations that are polynomial (or
even linear) in the number of input variables
\cite{bryant}. Consequently, ROBDDs have become indispensable tools in
the design and simulation of classical logic circuits.

\subsection{BDD Operations}
\label{sec:bdd_ops}

\begin{figure}[htb]
  \begin{center}
    \begin{tabular}{c|c|c}
      \hspace{-1mm}\includegraphics[width=5cm]{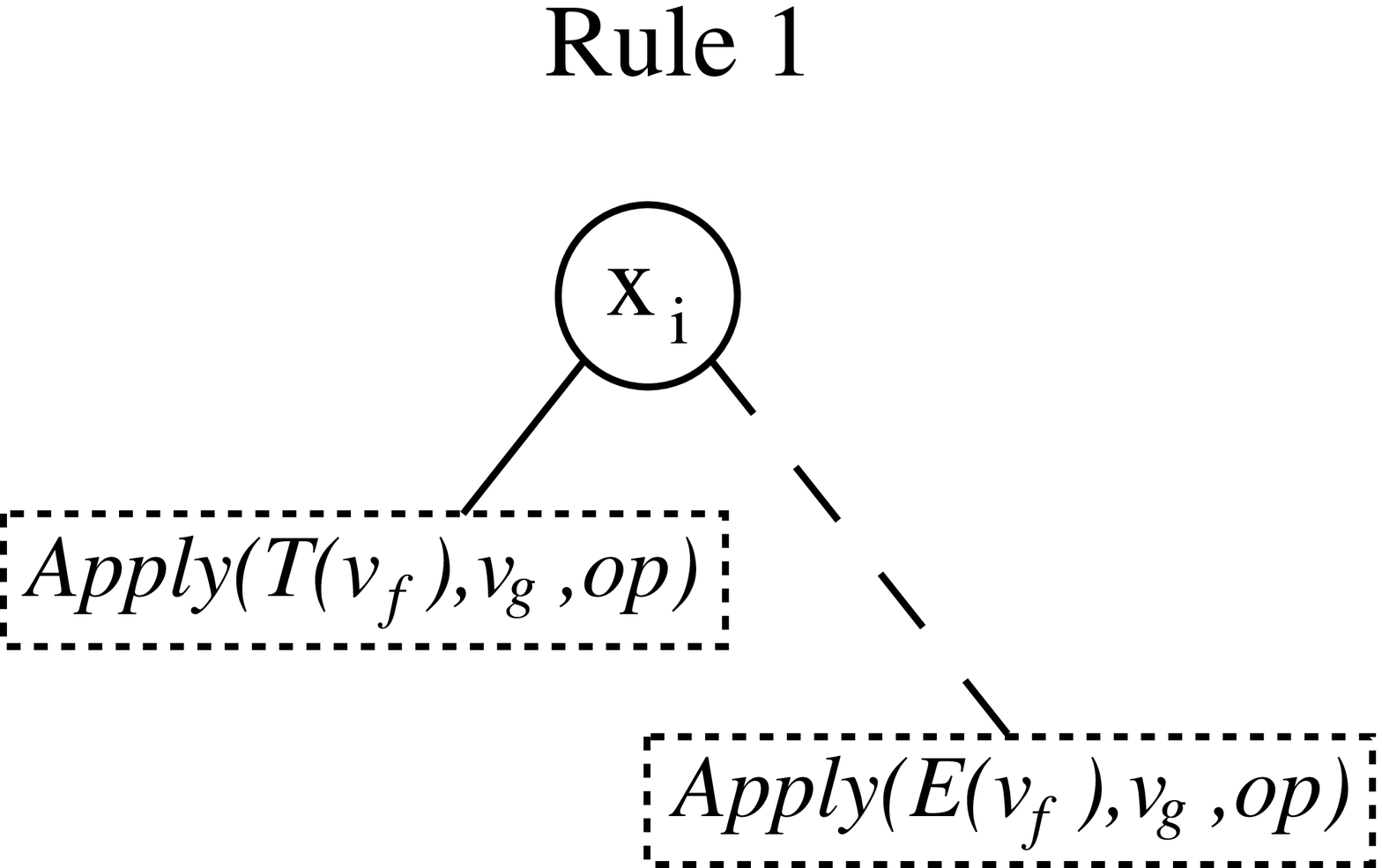}
      &
      \includegraphics[width=5cm]{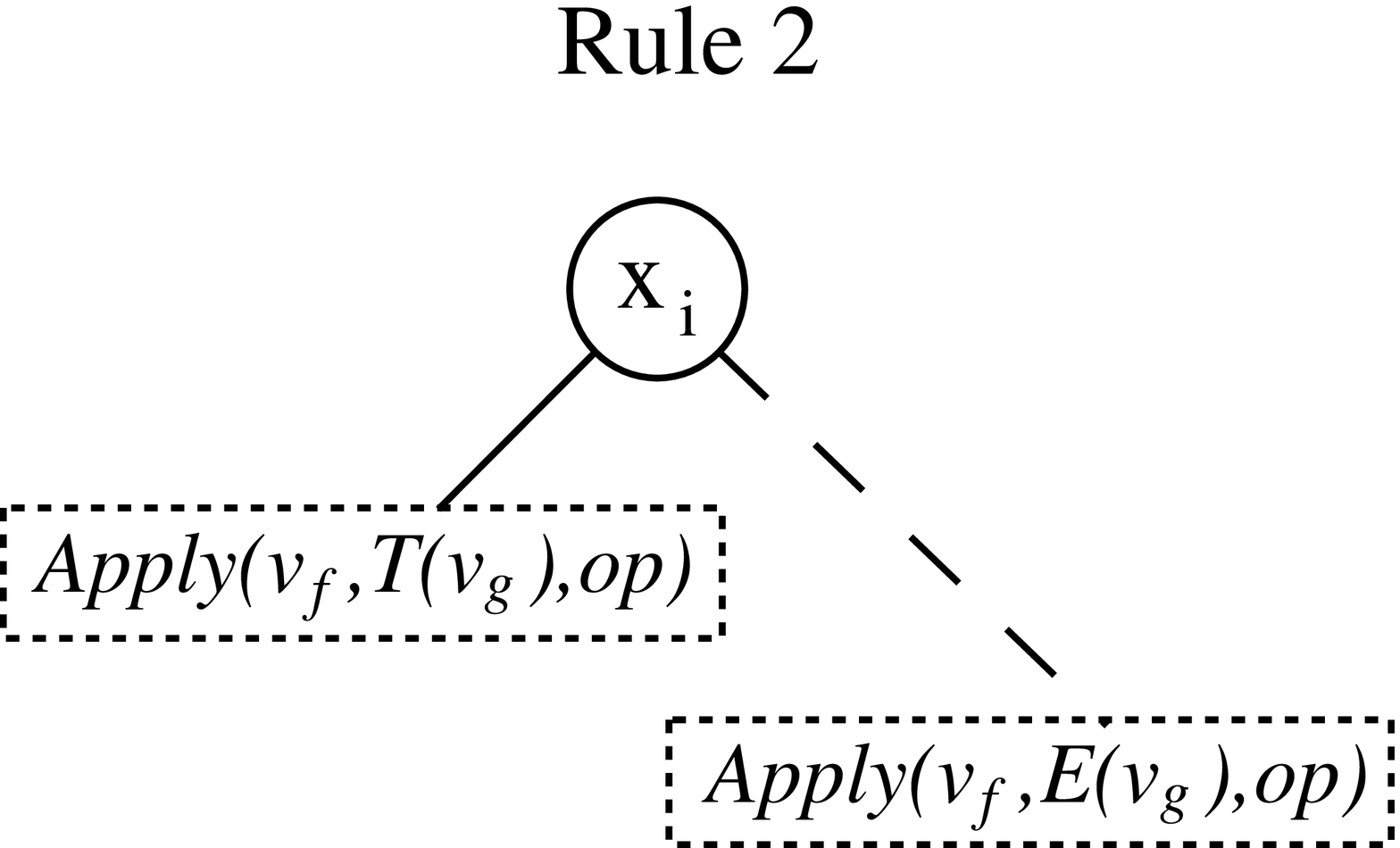}
      &
      \includegraphics[width=5.7cm]{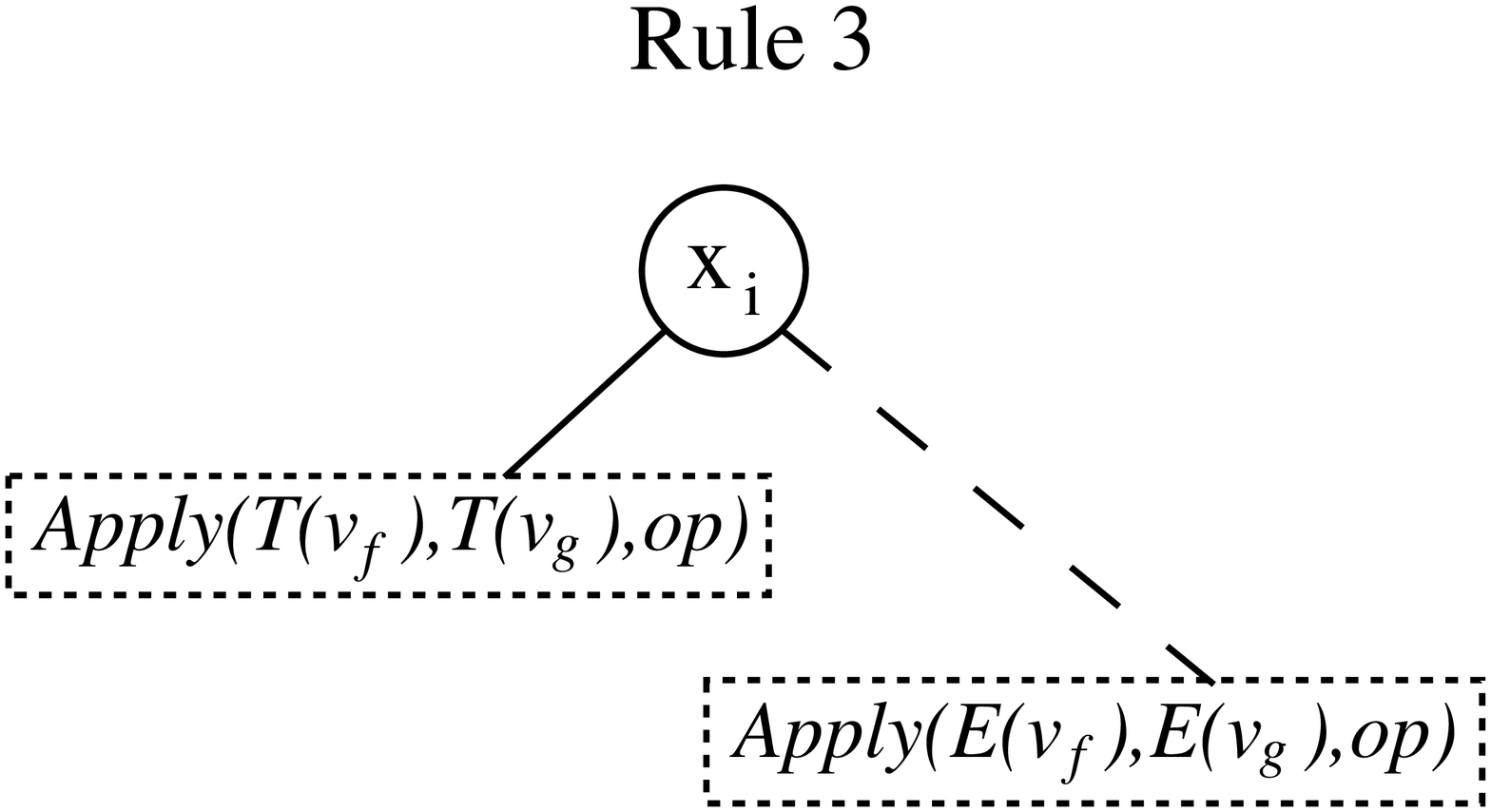}
      \\
      $x_i \prec x_j$
      &
      $x_i \succ x_j$
      &
      $x_i = x_j$
    \end{tabular}
    \parbox{17cm}{\caption{\label{fig:apply_rules} The three recursive
    rules used by the $Apply$ operation which determine how a new node
    should be added to a resultant ROBDD. In the figure, $x_i =
    Var(v_f)$ and $x_j = Var(v_g)$. The notation $x_i \prec x_j$ is
    defined to mean $x_i$ precedes $x_j$ in the variable
    ordering.}}
  \end{center}
\vspace{-2mm}
\end{figure}

Even though the ROBDD is often quite compact, efficient algorithms are
also needed to manipulate ROBDDs for circuit simulation. Thus, in
addition to the foregoing reduction rules, Bryant introduced a variety
of ROBDD operations with complexities that are bounded by the size of
the ROBDDs being manipulated \cite{bryant}. Of central importance is
the $Apply$ operation, which performs a binary operation with two
ROBDDs, producing a third ROBDD as the result. It can be used, for
example, to compute the logical $AND$ of two functions. $Apply$ is
implemented by a recursive traversal of the two ROBDD operands. For
each pair of nodes visited during the traversal, an internal node is
added to the resultant ROBDD using the three rules depicted in Fig.
\ref{fig:apply_rules}. To understand the rules, some notation must be
introduced. Let $v_f$ denote an arbitrary node in an ROBDD $f$.  If
$v_f$ is an internal node, $Var(v_f)$ is the Boolean variable
represented by $v_f$, $T(v_f)$ is the node reached when traversing the
{\em then} edge of $v_f$, and $E(v_f)$ is the node reached when
traversing the {\em else} edge of $v_f$.

Clearly the rules depend on the variable ordering. To illustrate,
consider performing $Apply$ using a binary operation $op$ and two
ROBDDs $f$ and $g$. $Apply$ takes as arguments two nodes, one from $f$
and one from $g$, and the operation $op$. This is denoted as
$Apply(v_f, v_g, op)$. $Apply$ compares $Var(v_f)$ and $Var(v_g)$ and
adds a new internal node to the ROBDD result using the three
rules. The rules also guide $Apply$'s traversal of the {\em then} and
{\em else} edges (this is the recursive step). For example, suppose
$Apply(v_f, v_g, op)$ is called and $Var(v_f) \prec Var(v_g)$. Rule 1
is invoked, causing an internal node containing $Var(v_f)$ to be added
to the resulting ROBDD. Rule 1 then directs the $Apply$ operation to
call itself recursively with $Apply(T(v_f), v_g, op)$ and
$Apply(E(v_f), v_g, op)$. Rules 2 and 3 dictate similar actions but
handle the cases when $Var(v_f) \succ Var(v_g)$ and $Var(v_f) =
Var(v_g)$. To recurse over both ROBDD operands correctly, the initial
call to $Apply$ must be $Apply(Root(f), Root(g), op)$ where $Root(f)$
and $Root(g)$ are the root nodes for the ROBDDs $f$ and $g$.

The recursion stops when both $v_f$ and $v_g$ are terminal nodes.
When this occurs, $op$ is performed with the values of the
terminals as operands, and the resulting value is added to the
ROBDD result as a terminal node. For example, if $v_f$ contains
the value logical $1$, $v_g$ contains the value logical $0$, and
op is defined to be $\oplus$ ($XOR$), then a new terminal with
value $1 \oplus 0 = 1$ is added to the ROBDD result. Terminal
nodes are considered {\em after} all variables are considered.
Thus, when a terminal node is compared to an internal node, either
Rule 1 or Rule 2 will be invoked depending on which ROBDD the
internal node is from.

The success of ROBDDs in making a seemingly difficult computational
problem tractable in practice led to the development of ROBDD variants
outside the domain of logic design. Of particular relevance to this
work are Multi-Terminal Binary Decision Diagrams (MTBDDs)
\cite{Clarke96} and Algebraic Decision Diagrams (ADDs)
\cite{Bahar97}. These data structures are compressed representations
of matrices and vectors rather than logic functions, and the amount of
compression achieved is proportional to the frequency of repeated
values in a given matrix or vector. Additionally, some standard
linear-algebraic operations, such as matrix multiplication, are
defined for MTBDDs and ADDs. Since they are based on the $Apply$
operation, the efficiency of these operations is proportional to the
size in nodes of the MTBDDs or ADDs being manipulated. Further
discussion of the MTBDD and ADD representations is deferred to Sec.
\ref{sec:sims} where the general structure of the QuIDD is described.

\subsection{Previous Simulation Techniques}
\label{sec:matrix_ops}

Quantum circuit simulators must support linear-algebraic operations
such as matrix multiplication, the tensor product, and the projection
operators. Simulation with the density matrix model additionally
requires the outer product and partial trace.\cite{Nielsen2000} Many
simulators typically employ array-based methods to facilitate these
operations and so require exponential computational resources in the
number of qubits. Such methods are often insensitive to the actual
values stored, and even sparse-matrix storage offers little
improvement for quantum operators with no zero matrix elements, such
as Hadamard operators. Previous work on these and other simulation
techniques are reviewed in this subsection.

One popular array-based simulation technique is to simulate $k$-input
quantum gates on an $n$-qubit state-vector ($k \leq n$) without
explicitly storing a $2^{n}\times 2^{n}$-matrix
representation.\cite{Obenland1997,qcsim} The basic idea is to simulate
the full-fledged matrix-vector multiplication by a series of simpler
operations. To illustrate, consider simulating a quantum circuit in
which a $1$-qubit Hadamard operator is applied to the third qubit of
the state-space $|00100 \rangle$. The state-vector representing this
state-space has $2^{5}$ elements. A naive way to apply the $1$-qubit
Hadamard is to construct a $2^{5}\times 2^{5}$ matrix of the form $I
\otimes I \otimes H \otimes I \otimes I$ and then multiply this matrix
by the state-vector. However, rather than compute $(I \otimes I
\otimes H \otimes I \otimes I)|00100 \rangle$, one can simply compute
$|00 \rangle \otimes H|1 \rangle \otimes |00 \rangle$, which produces
the same result using a $2\times 2$ matrix $H$. The same technique can
be applied when the state-space is in a superposition, such as $\alpha
|00100 \rangle + \beta |00000 \rangle$. In this case, to simulate the
application of a $1$-qubit Hadamard operator to the third qubit, one
can compute $|00 \rangle \otimes H(\alpha |1 \rangle + \beta |0
\rangle ) \otimes |00 \rangle$. As in the previous example, a $2\times
2$ matrix is sufficient.

While the above method allows one to compute a state space
symbolically, in a realistic simulation environment, state-vectors may
be much more complicated. Shortcuts that take advantage of the
linearity of matrix-vector multiplication are desirable. For example,
a single qubit can be manipulated in a state-vector by extracting a
certain set of two-dimensional vectors. Each vector in such a set is
composed of two probability amplitudes. The corresponding qubit states
for these amplitudes differ in value at the position of the qubit
being operated on but agree in every other qubit position. The
two-dimensional vectors are then multiplied by matrices representing
single qubit gates in the circuit being simulated. We refer to this
technique as {\em qubit-wise multiplication} because the state-space
is manipulated one qubit at a time. Obenland implemented a technique
of this kind as part of a simulator for quantum circuits
\cite{Obenland1997}.  His method applies one- and two-qubit operator
matrices to state vectors of size $2^n$. Unfortunately, in the best
case where $k = 1$, this only reduces the runtime and memory
complexity from $O(2^{2n})$ to $O(2^n)$, which is still exponential in
the number of qubits. 

Another implicit limitation of Obenland's implementation is that it
simulates with the state-vector representation only. The qubit-wise
technique has been extended, however, to enable density matrix
simulation by Black et al. and is implemented in NIST's QCSim
simulator.\cite{qcsim} As in its predecessor simulators, the arrays
representing density matrices in QCSim tend to grow exponentially. The
drawbacks of this asymptotic bottleneck are demonstrated
experimentally in Sec. \ref{sec:exp}.

Gottesman developed a simulation method involving the {\em Heisenberg
representation} of quantum computation which tracks the commutators of
operators applied by a quantum circuit \cite{Gottesman98}. With this
model, the state need not be represented explicitly by a state-vector
or a density matrix because the operators describe how an arbitrary
state-vector would be altered by the circuit. Gottesman showed that
simulation based on this model requires only polynomial memory and
runtime on a classical computer in certain cases. However, it appears
limited to the Clifford and Pauli groups of quantum operators, which
do not form a universal gate library.

Other advanced simulation techniques including MATLAB's ``packed''
representation, apply data compression to matrices and vectors, but
cannot perform matrix-vector multiplication without first
decompressing the matrices and vectors. A notable exception is Greve's
graph-based simulation of Shor's algorithm which uses BDDs
\cite{shornuf}.  Probability amplitudes of individual qubits are
modeled by single decision nodes.  Unfortunately, this only captures
superpositions where every participating qubit is rotated by $\pm 45$
degrees from $|0\rangle$ toward $|1\rangle$. Another BDD-based
technique proposed by Al-Rabadi et al. \cite{Alrabadi2002} can perform
multi-valued quantum logic. A drawback of this technique is that it is
limited to {\em synthesis} of quantum logic gates rather than
simulation of their behavior.

\section{GRAPH-BASED ALGORITHMS FOR DENSITY MATRIX SIMULATION}
\label{sec:sims}

QuIDDPro/D utilizes new simulation algorithms with unique features
that allow it to have much higher performance than naive explicit
array-based simulation techniques. These algorithms are the subject of
this section. In addition, we provide some implementation details
of the QuIDDPro/D simulator.

\subsection{QuIDDs and New QuIDD Algorithms}

Our density matrix simulation technique relies on the QuIDD data
structure. Previous work reported the use of QuIDDs in simulating the
state-vector model of quantum circuits \cite{aspdac,tech_report}. We
present new algorithms which use QuIDDs to efficiently perform the
outer product and the partial trace, both of which are needed to
simulate the density matrix representation. Before discussing the
details of these algorithms, we briefly review the QuIDD data
structure.

The QuIDD was born from the observation that vectors and matrices
which arise in quantum computing contain repeated structure. Operators
obtained from the tensor product of smaller matrices exhibit common
substructures which certain ROBDD variants can capture. The QuIDD can
be viewed as an ADD \cite{Bahar97} or MTBDD \cite{Clarke96} with
special properties \cite{aspdac,tech_report}.

\begin{figure}[!htb]
\begin{center}
  \includegraphics[width=9cm]{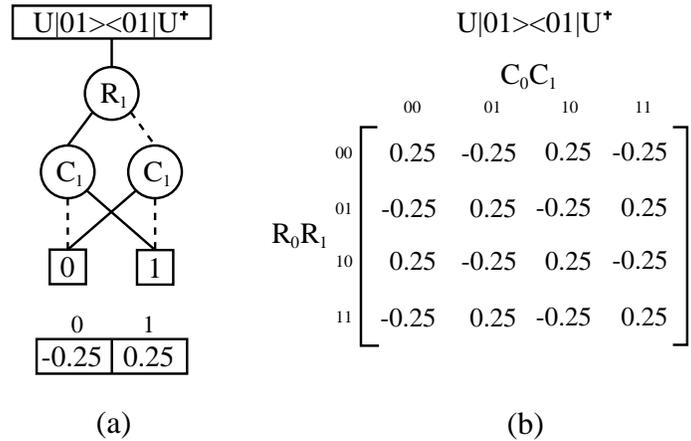}

  \parbox{16cm}{\caption{\label{fig:qdensity_eg} (a) QuIDD
  for the density matrix resulting from $U|01\rangle \langle
  01|U^{\dagger}$, where $U = H\otimes H$, and (b) its explicit matrix
  form.}}
\end{center}
\vspace{-2mm}
\end{figure}

Figure \ref{fig:qdensity_eg}a shows the QuIDD that results from
applying $U$ to an outer product as $U|01\rangle \langle
01|U^{\dagger}$, where $U = H\otimes H$. The $R_i$ nodes of the QuIDD
encode the binary indices of the rows in the explicit
matrix. Similarly, the $C_i$ nodes encode the binary indices of the
columns. Solid lines leaving a node denote the positive cofactor of
the index bit variable (a value of $1$), while dashed lines denote the
negative cofactor (a value of $0$). Terminal nodes correspond to the
value of the element in the explicit matrix whose binary row/column
indices are encoded by the path that was traversed.

Notice that the first and second pairs of rows of the explicit matrix
in Fig. \ref{fig:qdensity_eg}b are equal, as are the first and
second pairs of columns. This redundancy is captured by the QuIDD in
Fig. \ref{fig:qdensity_eg}a because the QuIDD does not contain any
$R_0$ or $C_0$ nodes. In other words, the values and their locations
in the explicit matrix can be completely determined without the
superfluous knowledge of the first row and column index bits.

Measurement, matrix multiplication, addition, scalar products, the
tensor product, and other operations involving QuIDDs are variations
of the well-known $Apply$ algorithm discussed in
Sec. \ref{sec:bdd_ops}.\cite{aspdac,tech_report} It has been proven
that by interleaving the row and column variables in the variable
ordering, QuIDDs can represent and operate on a certain class of
matrices and vectors using time and memory resources that are {\em
polynomial} in the number of qubits. This class includes, but is not
limited to, any equal superposition of $n$ qubits, any sequence of $n$
qubits in the computational basis states, $n$-qubit Pauli operators,
and $n$-qubit Hadamard operators \cite{tech_report}.

Since QuIDDs already have the capability to represent matrices and
multiply them \cite{aspdac,tech_report}, extending QuIDDs to
encompass the density matrix requires algorithms for the outer product
and the partial trace. The outer product involves matrix
multiplication between a column vector and its complex-conjugate
transpose. Since a column vector QuIDD only depends on row variables,
the transpose can be accomplished by swapping the row variables with
column variables. The complex conjugate can then be performed with a
DFS traversal that replaces terminal node values with their complex
conjugates. The original column vector QuIDD is then multiplied by its
complex-conjugate transpose using the matrix multiply operation
previously defined for QuIDDs \cite{aspdac,tech_report}. Pseudo-code
for this algorithm is given in Fig. \ref{fig:pseudo_outer}. Notice
that before the result is returned, it is divided by
$2^{num\_qubits}$, where $num\_qubits$ is the number of qubits
represented by the QuIDD vector. This is done because a QuIDD that
only depends on $n$ row variables can be viewed as either a $2^n
\times 1$ column vector or a $2^n \times 2^n$ matrix in which all
columns are the same. Since matrix multiplication is performed in
terms of the latter case \cite{aspdac,tech_report,Bahar97}, the
result of the outer product contains values that are multiplied by an
extra factor of $2^n$, which must be normalized.

\begin{figure}[!tb]
\begin{center}
  \begin{tabular}{c|c}
    \parbox{7cm}{
      \vspace{-20mm}
      $\begin{array}{l}
	\mathbf{Outer\_Product} (Q, num_qubits)\ \{ \\
	\hspace{1em}Q\_cctrans = \mathbf{Swap\_Row\_Col\_Variables} (Q); \\
	\hspace{1em}Q\_cctrans = \mathbf{Complex\_Conj} (Q\_cctrans); \\
	\hspace{1em}R = \mathbf{Matrix\_Multiply} (Q, Q\_cctrans); \\
	\hspace{1em}R = \mathbf{Scalar\_Div} (Q\_cctrans, 2^{num\_qubits}); \\
	\hspace{1em}return\ R; \\
	\}
      \end{array}
      $
    }
    &
    \parbox{7cm}{
      $\begin{array}{l}
	\mathbf{Complex\_Conj} (Q)\ \{ \\
	\hspace{1em}if\ (\mathbf{Is\_Constant} (Q)) \\
	\hspace{2em}return\ \mathbf{New\_Terminal} ({\mathbf Real} (Q), -1*{\mathbf Imag} (Q)); \\
	\hspace{1em}if\ (\mathbf{Table\_Lookup} (computed\_table, Q, R) \\
	\hspace{2em}return R; \\
	\hspace{1em}v = \mathbf{Top\_Var} (Q); \\
	\hspace{1em}T = \mathbf{Complex\_Conj} (Q_{v}); \\
	\hspace{1em}E = \mathbf{Complex\_Conj} (Q_{v'}); \\
	\hspace{1em}R = \mathbf{ITE} (v, T, E); \\
	\hspace{1em}\mathbf{Table\_Insert} (computed\_table, Q, R); \\
	\hspace{1em} return\ R; \\
	\}  
      \end{array}
      $
    }
    \\
    (a) & (b)
  \end{tabular}
  \parbox{16cm}{\caption{\label{fig:pseudo_outer}Pseudo-code for (a)
  the QuIDD outer product and (b) its complex conjugation helper
  function ${\mathbf Complex\_Conj}$. The code for ${\mathbf
  Scalar\_Div}$ is the same as ${\mathbf Complex\_Conj}$, except that
  in the terminal node case it returns the value of the terminal
  divided by a scalar. Other functions are typical ADD operations.\cite{Bahar97,cudd}}}  %\cite{Bahar97, cudd} }}

\end{center}
\vspace{-4mm}
\end{figure}

To motivate the QuIDD-based partial trace algorithm, we note how the
partial trace can be performed with explicit matrices. The trace of a
matrix $A$ is the sum of $A$'s diagonal elements. To perform the
partial trace over a particular qubit in an $n$-qubit density matrix,
the trace operation can be applied iteratively to sub-matrices of the
density matrix. Each sub-matrix is composed of four elements with row
indices $r0s$ and $r1s$, and column indices $c0d$and $c1d$, where $r$,
$s$, $c$, and $d$ are arbitrary sequences of bits which index the
$n$-qubit density matrix.

Tracing over these sub-matrices has the effect of reducing the
dimensionality of the density matrix by one qubit. A well-known ADD
operation which reduces the dimensionality of a matrix is the
$Abstract$ operation \cite{Bahar97}. Given an arbitrary ADD $f$,
abstraction of variable $x_i$ eliminates $x_i$ from the support of $f$
by combining the positive and negative cofactors of $f$ with respect
to $x_i$ using some binary operation. In other words, $Abstract(f,
x_i, op) = f_{x_i}\ op\ f_{x_i'}$.

\begin{figure}[!tb]
\begin{center}
  $\begin{array}{l}
    \mathbf{Ptrace} (Q, qubit\_index)\ \{ \\
    \hspace{1em}if (\mathbf{Is\_Constant} (Q)) \\
    \hspace{2em}return\ Q; \\
    \hspace{1em}top\_q = \mathbf{Top\_Var}
    \hspace{1em}if\ (qubit\_index < \mathbf{Index} (top\_q))\ \{ \\
    \hspace{2em}R = \mathbf{Apply} (Q, Q, +); \\
    \hspace{2em}return\ R; \\
    \hspace{1em}\} \\ \\
    \hspace{1em}if\ (\mathbf{Table\_Lookup} (computed\_table, (Q, qubit\_index), R) \\
    \hspace{2em}return\ R; \\
    \hspace{1em}T = Q_{top\_q}; \\
    \hspace{1em}E = Q_{top\_q'}; \\ \\
    \hspace{1em}if\ (qubit\_index == \mathbf{Index} (top\_q))\ \{ \\
    \hspace{2em}if\ (\mathbf{Is\_Constant} (T)\ ||\ \mathbf{Index} (T) > \mathbf{Index} (Q) + 1) \\
    \hspace{3em}r_1 = \mathbf{Ptrace} (T, qubit\_index); \\
    \hspace{2em}else\ \{ \\
    \hspace{3em}top\_T = \mathbf{Top\_Var} (T); \\
    \hspace{3em}r_1 = \mathbf{Ptrace} (T_{top\_T}, qubit\_index); \\
    \hspace{2em}\} \\ \\
    \hspace{2em}if\ (\mathbf{Is\_Constant} (E)\ ||\ \mathbf{Index} (E) > \mathbf{Index} (Q) + 1) \\
    \hspace{3em}r_2 = \mathbf{Ptrace} (E, qubit\_index); \\
    \hspace{2em}else\ \{ \\
    \hspace{3em}top\_E \ \mathbf{Top\_Var} (E); \\
    \hspace{3em}r_2 = \mathbf{Ptrace} (E_{top\_E'}, qubit\_index); \\
    \hspace{2em}\} \\
    \hspace{2em}R = \mathbf{Apply} (r_1, r_2, +); \\
    \hspace{2em}\mathbf{Table\_Insert} (computed\_table, (Q, qubit\_index), R); \\
    \hspace{2em}return\ R; \\
    \hspace{1em}\} \\ \\
    \hspace{1em}else\ \{\hspace{1em}/*\ (qubit\_index > \mathbf{Index} (top\_q)\ */ \\
    \hspace{2em}r_1 = \mathbf{Ptrace} (T, qubit\_index); \\
    \hspace{2em}r_2 = \mathbf{Ptrace} (E, qubit\_index); \\
    \hspace{2em}R = \mathbf{ITE} (top\_q, r_1, r_2); \\
    \hspace{2em}\mathbf{Table\_Insert} (computed\_table, (Q, qubit\_index), R); \\
    \hspace{2em}return\ R; \\
    \hspace{1em}\} \\
    \}
    \end{array}
  $

  \parbox{16cm}{\caption{\label{fig:pseudo_ptrace} Pseudo-code
  for the QuIDD partial trace. The index of the qubit being traced
  over is $qubit\_index$. Compare this to the pseudo-code for
  $Abstract()$.\cite{Bahar97}}} 
\end{center}
\vspace{-3mm}
\end{figure}

For QuIDDs, there is a one-to-one correspondence between a qubit on
wire $i$ (wires are labeled top-down starting at $0$) and variables
$R_i$ and $C_i$. So at first glance, one may suspect that the partial
trace of qubit $i$ in $f$ can be achieved by a performing $Abstract(f,
R_i, +)$ followed by $Abstract(f, C_i, +)$. However, this will add the
rows determined by qubit $i$ independently of the columns. The desired
behavior is to perform the diagonal addition of sub-matrices by
accounting for both the row and column variables due to $i$ {\em
simultaneously}. The pseudo-code to perform the partial trace
correctly is depicted in Fig. \ref{fig:pseudo_ptrace}. In comparing
this with the pseudo-code for the $Abstract$ algorithm \cite{Bahar97},
the main difference is that when $R_i$ corresponding to qubit $i$ is
reached, we take the positive and negative cofactors {\em twice}
before making the recursive call. Since the interleaved variable
ordering of QuIDDs guarantees that $C_i$ immediately follows $R_i$
\cite{aspdac,tech_report}, taking the positive and negative cofactors
twice simultaneously abstracts both the row and column variables for
qubit $i$, achieving the desired result of summing diagonals. In other
words, for a QuIDD $f$, the partial trace over qubit $i$ is $Ptrace(f,
i) = f_{R_iC_i} + f_{R_i'C_i'}$. Note that in the pseudo-code there
are checks for the special case when $C_i$ is not in the support of
the QuIDD. Not shown in the pseudo-code is book-keeping which shifts
up the variables in the resulting QuIDD to fill the hole in the
ordering left by the row and column variables that were traced-over.

\subsection{QuIDDPro/D}

QuIDDPro/D is the implementation of our simulation technique. It was
written in C++, and the source code is approximately $17,000$ lines
long. The density matrix is represented by a QuIDD class with terminal
node values of type $complex<long\ double>$. Gate operators are also
represented as QuIDDs. QuIDDPro/D utilizes our earlier QuIDDPro source
code \cite{aspdac,tech_report} and extends it significantly with an
implementation of the outer product and partial trace pseudo-code of
Fig. \ref{fig:pseudo_outer} and
Fig. \ref{fig:pseudo_ptrace}. Additionally, the technique of using an
epsilon to deal with precision problems in QuIDDPro
\cite{aspdac,tech_report} has been replaced with a technique that
rounds complex numbers to $25$ significant digits. This enhancement
allows an end-user to avoid having to find an optimal value of epsilon
for a given quantum circuit input. A front-end parser was also created
using Flex and Bison to accept a subset of the MATLAB language, which
is ideal for describing linear-algebraic operations in a text
format. In addition to incorporating a set of well-known numerical
functions, the language also supports a number of other functions that
are useful in quantum circuit simulation. For example, there is a
function for creating controlled-$U$ gates with an arbitrary
configuration for the control qubits and user-defined specification of
$U$. Functions to perform deterministic measurement, probabilistic
measurement, and the partial trace, among others, are also
supported. The current version of QuIDD\-Pro/D contains over $65$
functions and language features.

\section{EXPERIMENTAL RESULTS}
\label{sec:exp}

We consider a number of quantum circuit benchmarks which cover errors,
error correction, reversible logic, communication, and quantum
search. We devised some of the benchmarks, while others are drawn from
NIST\cite{qcsim} and from a site devoted to reversible
circuits\cite{maslov}. For every benchmark, the simulation performance
of QuIDD\-Pro/D is compared with NIST's QCSim quantum circuit
simulator, which utilizes an explicit array-based computational
engine. The results indicate that QuIDD\-Pro/D far outperforms
QCSim. All experiments are performed on a 1.2GHz AMD Athlon
workstation with 1GB of RAM running Linux.

\subsection{Reversible Circuits}

Here we examine the performance of QuIDD\-Pro/D simulating a set of
reversible circuits, which we define as quantum circuits that perform
classical operations.\cite{Nielsen2000} Specifically, if the input
qubits of a quantum circuit are all in the computational basis
(i.e. they have only $|0 \rangle$ or $|1 \rangle$ values), there is no
quantum noise, and all the gates are NOT variants such as CNOT,
Toffoli, X, etc, then the output qubits and all intermediate states
will also be in the computational basis. Such a circuit results in a
classical logic operation which is reversible in the sense that the
inputs can always be derived from the outputs and the circuit
function. Reversibility comes from the fact that all quantum operators
must be unitary and therefore all have inverses.\cite{Nielsen2000}

\begin{figure}[tb]
  \begin{center}
    \begin{tabular}{c|c}
      \includegraphics[width=4.7cm,height=3.7cm]{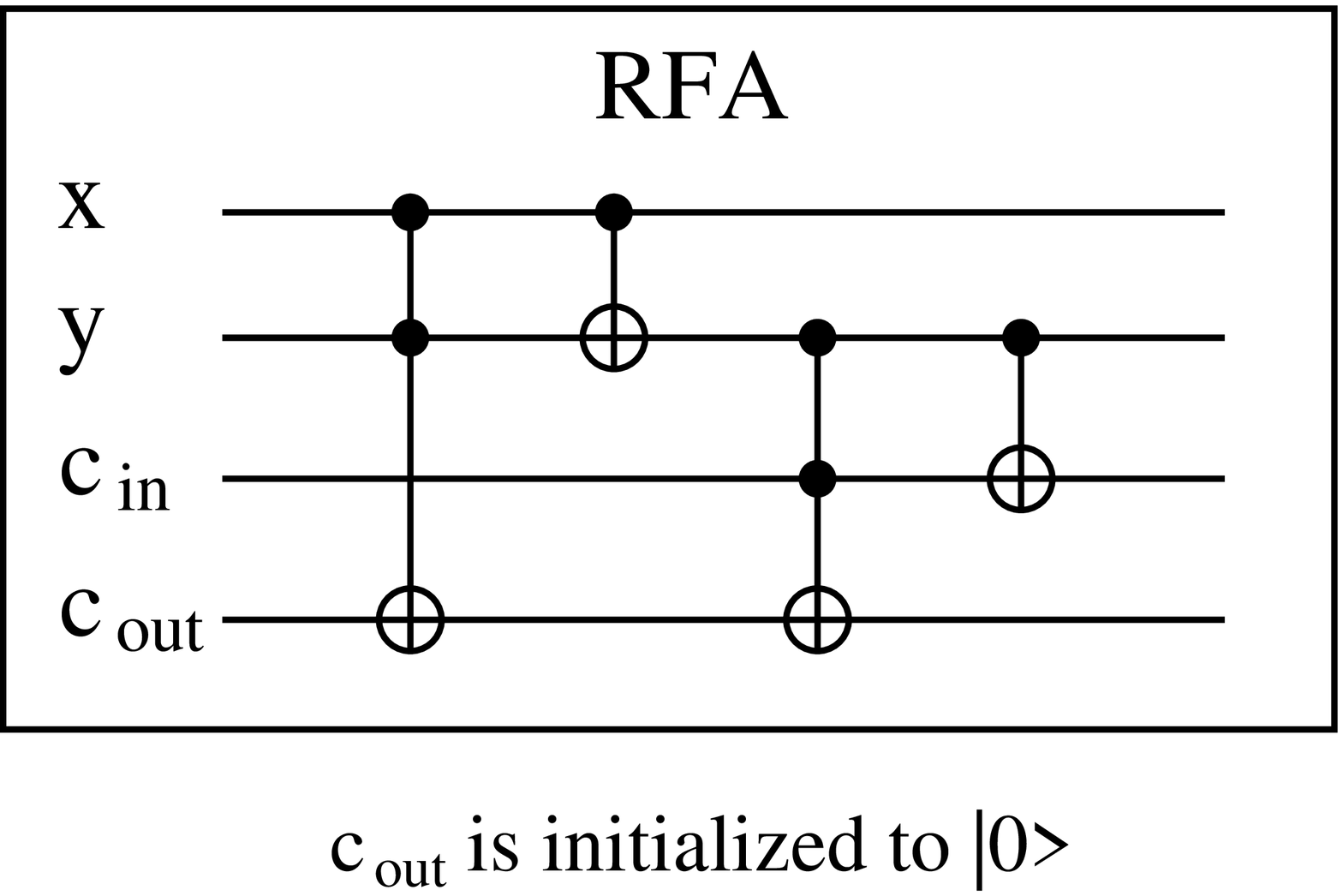}
      &
      \includegraphics[width=10.5cm]{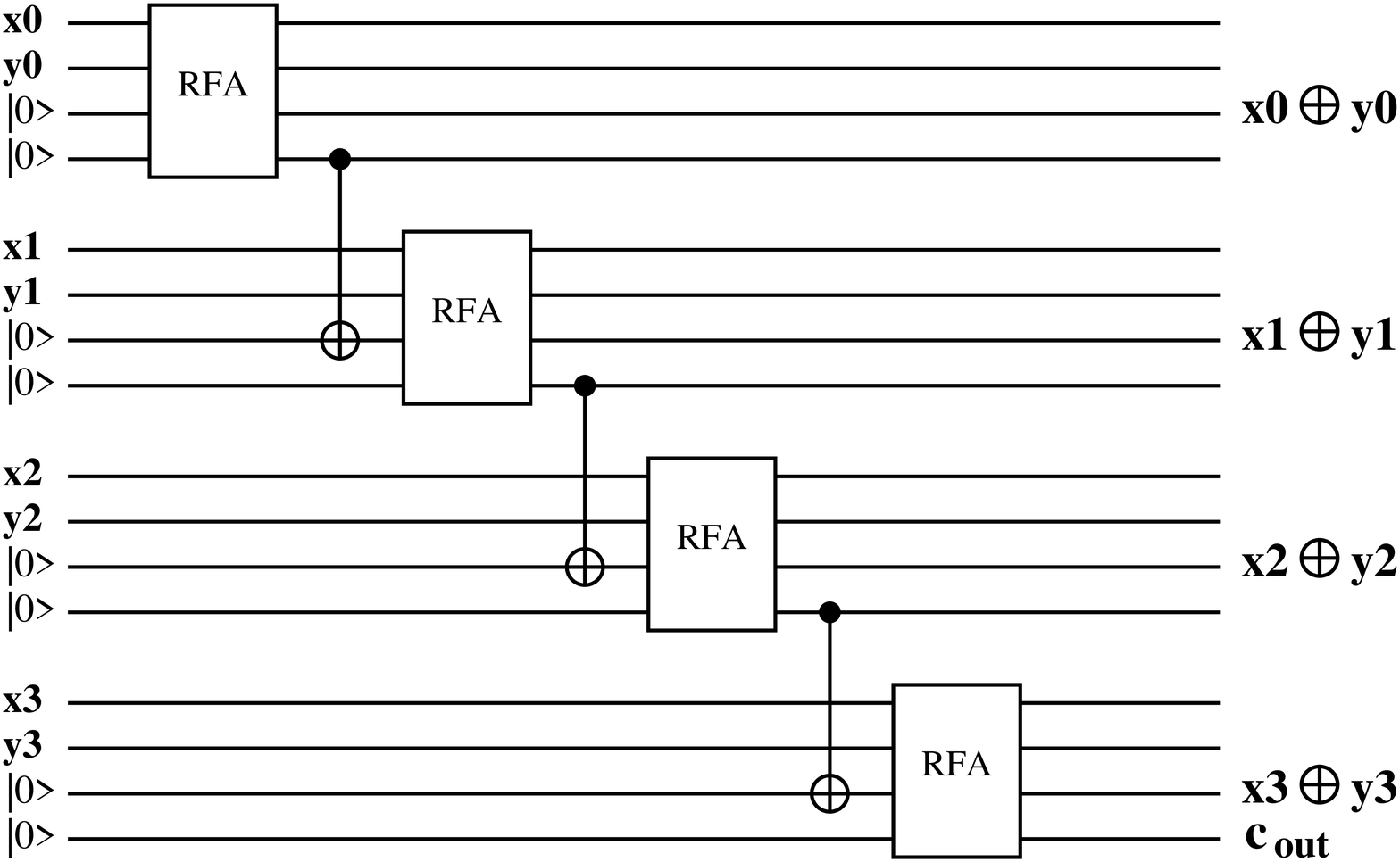} \\
      (a) & (b)
    \end{tabular}
    \parbox{16cm}{\caption{\label{fig:rc_adder} (a) An implementation
	of a reversible full-adder (RFA), and (b) a reversible $4$-bit
	ripple-carry adder which utilizes the RFA as a module. The
	reversible ripple-carry adder circuit computes the binary sum
	of two $4$-bit numbers: $x_3x_2x_1x_0 \oplus
	y_3y_2y_1y_0$. $c_{out}$ is the final carry bit output from
	the addition of the most-significant bits ($x_3$ and
	$y_3$). }}
  \end{center}
  \vspace{-4mm}
\end{figure}

The first reversible benchmark we consider is a reversible $4$-bit
ripple-carry adder which is depicted in Fig.
\ref{fig:rc_adder}. Since the size of a QuIDD is sensitive to the
arrangement of different values of matrix elements, we simulate the
adder with varied input values (``rc\_adder1'' through
``rc\_adder4''). This is also done for other benchmarks.

Two other reversible benchmarks we simulate contain fewer qubits but
more gates than the ripple-carry adder. One of these benchmarks is a
$12$-qubit reversible circuit that outputs a $|1 \rangle$ on the last
qubit if and only if the number of $|1 \rangle$'s in the input qubits
is $3$, $4$, $5$, or $6$ (``9sym1'' through ``9sym5'').\cite{maslov}
The other benchmark is a $15$-qubit reversible circuit that generates
the classical Hamming code of the input qubits (``ham15\_1'' through
``ham15\_3'').\cite{maslov}

Performance results for all of these benchmarks are shown in Tab.
\ref{tab:rev_results}. QuIDD\-Pro/D significantly outperforms QCSim in
every case. In fact for circuits of $14$ or more qubits, QCSim
requires more than 2GB of memory.  Since QCSim uses an explicit
array-based engine, it is insensitive to the arrangement and values of
elements in matrices. Therefore, one can expect QCSim to use more than
2GB of memory for {\em any} benchmark with $14$ or more qubits,
regardless of the circuit functionality and input values. Another
interesting result is that even though QuIDD\-Pro/D is, in general,
sensitive to the arrangement and values of matrix elements, the data
indicate that QuIDD\-Pro/D is insensitive to varied inputs on the same
circuit for error-free reversible benchmarks. However, QuIDD\-Pro/D
still compresses the tremendous amount of redundancy present in these
benchmarks, and its performance does vary from one type of circuit to
the next.

\begin{table}[!tb]
\begin{center}
    \parbox{16cm}{
    \caption{\label{tab:rev_results} Performance results for QuIDDPro/D and
    QCSim on the reversible circuit benchmarks. MEM-OUT indicates
    that a memory usage cutoff of 2GB was exceeded.}
    \vspace{2mm}
    }
  \begin{tabular}{|c|c|c||c|c||c|c|} \hline
    Benchmark & No. of & No. of & \multicolumn{2}{|c||}{QuIDDPro/D} & \multicolumn{2}{|c|}{QCSim} \\ \cline{4-7}
    & Qubits & Gates & Runtime (s) & Peak Memory (MB) & Runtime (s) & Peak Memory (MB) \\ \hline
    rc\_adder1 & 16 & 24 & 0.44 & 0.0625 & MEM-OUT & MEM-OUT \\ \hline
    rc\_adder2 & 16 & 24 & 0.44 & 0.0625 & MEM-OUT & MEM-OUT \\ \hline
    rc\_adder3 & 16 & 24 & 0.44 & 0.0625 & MEM-OUT & MEM-OUT \\ \hline
    rc\_adder4 & 16 & 24 & 0.44 & 0.0625 & MEM-OUT & MEM-OUT \\ \hline
    9sym1 & 12 & 29 & 0.2 & 0.0586 & 8.01 & 128.1 \\ \hline
    9sym2 & 12 & 29 & 0.2 & 0.0586 & 8.02 & 128.1 \\ \hline
    9sym3 & 12 & 29 & 0.2 & 0.0586 & 8.04 & 128.1 \\ \hline
    9sym4 & 12 & 29 & 0.2 & 0.0586 & 8 & 128.1 \\ \hline
    9sym5 & 12 & 29 & 0.2 & 0.0586 & 7.95 & 128.1 \\ \hline
    ham15\_1 & 15 & 148 & 1.99 & 0.121 & MEM-OUT & MEM-OUT \\ \hline
    ham15\_2 & 15 & 148 & 2.01 & 0.121 & MEM-OUT & MEM-OUT \\ \hline
    ham15\_3 & 15 & 148 & 1.99 & 0.121 & MEM-OUT & MEM-OUT \\ \hline
  \end{tabular}
\end{center}
\vspace{-6mm}
\end{table}

\subsection{Error Correction and Communication}

Now we analyze the performance of QuIDD\-Pro/D on simulations that
incorporate errors and error correction. We consider some simple
benchmarks that encode single qubits into Steane's $7$-qubit
error-correcting code\cite{Steane1996,Nielsen2000} and some more
complex benchmarks that use the Steane code to correct a combination
of bit-flip and phase-flip errors in a half-adder and Grover's quantum
search algorithm.\cite{Grover97} Secure quantum communication is also
considered here because eavesdropping disrupts a quantum channel and
can be treated as an error.

The first set of benchmarks, ``steaneX'' and ``steaneZ,'' each encode
a single logical qubit as seven physical qubits with the Steane code
and simulate the effect of a probabilistic bit-flip and phase-flip
error, respectively.\cite{qcsim} ``steaneZ'' contains $13$ qubits
which are initialized to the mixed state $0.866025|0000000000000
\rangle \\+ 0.5|0000001000000 \rangle$. A combination of gates apply a
probabilistic phase-flip on one of the qubits and calculate the error
syndrome and error rate. ``steaneX'' is a $12$ qubit version of the
same circuit that simulates a probabilistic bit-flip error.

A more complex benchmark that we simulate is a reversible half-adder
with three logical qubits that are encoded into twenty one physical
qubits with the Steane code. Additionally, three ancillary qubits are
used to track the error rate, giving a total circuit size of twenty
four qubits. ``hadder1\_bf1'' through ``hadder3\_bf3'' simulate the
half-adder with different numbers of bit-flip errors on various
physical qubits in the encoding of one of the logical qubit
inputs. Similarly, ``hadder1\_pf1'' through ``hadder3\_pf3'' simulate
the half-adder with various phase-flip errors.  

Another large benchmark we simulate is an instance of Grover's quantum
search algorithm. Grover's algorithm searches for a subset of items in
an unordered database of $N$ items. Allowed selection criteria are
black-box predicates, called oracles, that can be evaluated on any
database record. This particular benchmark applies an oracle that
searches for one element in a database of four items. It has two
logical data qubits and one logical oracle ancillary qubit which are
all encoded with the Steane code. Like the half-adder circuit, this
results in a total circuit size of twenty four
qubits. ``grover\_s1'' simulates the circuit with the encoded
qubits in the absence of errors. ``grover\_s\_bf1'' and
``grover\_s\_pf1'' introduce and correct a bit-flip and
phase-flip error, respectively, on one of the physical qubits in the
encoding of the logical oracle qubit.

\begin{figure}[tb]
  \vspace{2mm}
  \begin{center}
    \hspace{-2mm}\includegraphics[width=15cm]{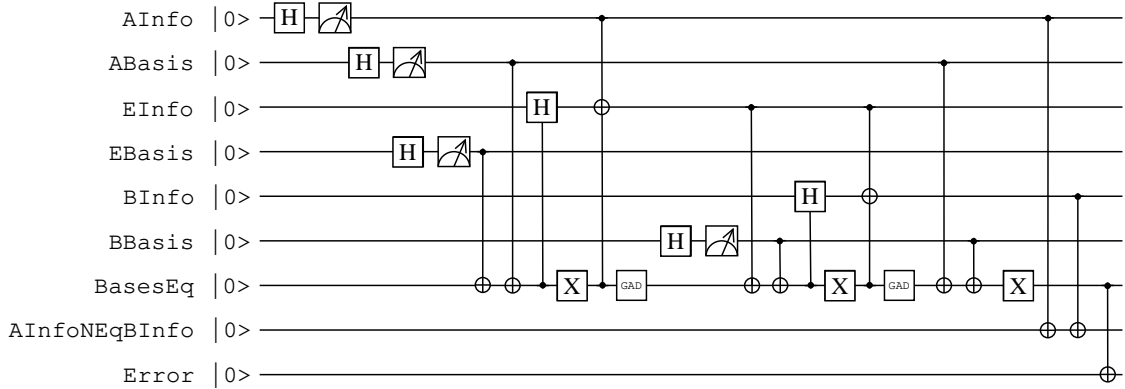}
    \hspace{-3mm}\parbox{17cm}{\caption{\label{fig:bb84} Quantum
    circuit for the ``bb84Eve'' benchmark.}} 
  \end{center}
  \vspace{-5mm}
\end{figure}

In addition to error modeling and error correction for computational
circuits, another important application is secure communication using
quantum cryptography. The basic concept is to use a quantum-mechanical
phenomenon called entanglement to distribute a shared
key. Eavesdropping constitutes a measurement of the quantum state
representing the key, disrupting the quantum state. This disruption
can be detected by the legitimate communicating parties. Since actual
implementations of quantum key distribution have already been
demonstrated \cite{magiq}, efficient simulation of these protocols may
play a key role in exploring possible improvements. Therefore, we
present two benchmarks which implement BB84, one of the earliest
quantum key distribution protocols \cite{bb84}. ``bb84Eve'' accounts
for the case in which an eavesdropper is present (see Fig.
\ref{fig:bb84}) and contains $9$ qubits, whereas ``bb84NoEve''
accounts for the case in which no eavesdropper is present and contains
$7$ qubits. In both circuits, all qubits are traced-over at the end
except for two qubits reserved to track whether or not the legitimate
communicating parties successfully shared a key (BasesEq) and the
error due to eavesdropping (Error).

Performance results for all of these benchmarks are show in Tab.
\ref{tab:err_results}. Again, QuIDD\-Pro/D significantly outperforms
QCSim on all benchmarks except for ``bb84Eve'' and ``bb84NoEve.''  The
performance of QuIDD\-Pro/D and QCSim is about the same for these
benchmarks.  The reason is that these benchmarks contain fewer qubits
than all of the others. Since each additional qubit doubles the size
of an explicit density matrix, QCSim has difficulty simulating the
larger Steane encoded benchmarks.

\begin{table}[!tb]
\begin{center}
  \parbox{16cm}{
    \caption{\label{tab:err_results} Performance results for QCSim and
      QuIDDPro/D on the error-related benchmarks. Runtime results for ``bb84NoEve'' are below reasonable measurement error. MEM-OUT indicates that a memory usage cutoff of 2GB was exceeded.}  
    \vspace{2mm}
  }
  \begin{tabular}{|c|c|c||c|c||c|c|} \hline
    Benchmark & No. of  & No. of & \multicolumn{2}{|c||}{QuIDDPro/D} & \multicolumn{2}{|c|}{QCSim} \\ \cline{4-7}
    & Qubits & Gates & Runtime (s) & Peak Memory (MB) & Runtime (s) & Peak Memory (MB) \\ \hline
    steaneZ & 13 & 143 & 0.6 & 0.672 & 287 & 512 \\ \hline
    steaneX & 12 & 120 & 0.27 & 0.68 & 53.2 & 128 \\ \hline
    hadder\_bf1 & 24 & 49 & 18.3 & 1.48 & MEM-OUT & MEM-OUT \\ \hline
    hadder\_bf2 & 24 & 49 & 18.7 & 1.48 & MEM-OUT & MEM-OUT \\ \hline
    hadder\_bf3 & 24 & 49 & 18.7 & 1.48 & MEM-OUT & MEM-OUT \\ \hline
    hadder\_pf1 & 24 & 51 & 21.2 & 1.50 & MEM-OUT & MEM-OUT \\ \hline
    hadder\_pf2 & 24 & 51 & 21.2 & 1.50 & MEM-OUT & MEM-OUT \\ \hline
    hadder\_pf3 & 24 & 51 & 20.7 & 1.50 & MEM-OUT & MEM-OUT \\ \hline
    grover\_s1 & 24 & 50 & 2301 & 94.2 & MEM-OUT & MEM-OUT \\ \hline
    grover\_s\_bf1 & 24 & 71 & 2208 & 94.3 & MEM-OUT & MEM-OUT \\ \hline
    grover\_s\_pf1 & 24 & 73 & 2258 & 94.2 & MEM-OUT & MEM-OUT \\ \hline
    bb84Eve & 9 & 26 & 0.02 & 0.129 & 0.19 & 2.0 \\ \hline
    bb84NoEve & 7 & 14 & $<$0.01 & 0.0313 & $<$0.01 & 0.152 \\ \hline
  \end{tabular}
\end{center}
\vspace{-4mm}
\end{table}

\subsection{Scalability and Quantum Search}

To test scalability with the number of input qubits, we turn to
quantum circuits containing a variable number of input qubits. In
particular, we reconsider Grover's quantum search algorithm. However,
for these instances of quantum search, the qubits are not encoded with
the Steane code, and errors are not introduced. The oracle performs
the same function as the one described in the last subsection except
that the number of data qubits ranges from five to twenty. 

Performance results for these circuit benchmarks are shown in Tab.
\ref{tab:grover}. Again, QuIDDPro/D has significantly better
performance. These results highlight the fact that QCSim's explicit
representation of the density matrix becomes an asymptotic bottleneck
as $n$ increases, whereas QuIDDPro/D's compression of the density
matrix and operators scales extremely well.

\begin{table}[!tb]
%\small
\begin{center}
  \parbox{15cm}{
  \caption{\label{tab:grover} Performance results for QCSim and QuIDDPro/D on the Grover's quantum search benchmark. MEM-OUT indicates that a memory usage cutoff of 2GB was exceeded.}
  \vspace{2mm}
  }
  \begin{tabular}{|c|c||c|c||c|c|} \hline
    No. of & No. of & \multicolumn{2}{|c||}{QuIDDPro/D} & \multicolumn{2}{|c|}{QCSim} \\ \cline{3-6}
    Qubits & Gates & Runtime (s) & Peak Memory (MB) & Runtime (s) & Peak Memory (MB) \\ \hline
    5 & 32 & 0.05 & 0.0234 & 0.01 & 0.00781 \\ \hline
    6 & 50 & 0.07 & 0.0391 & 0.01 & 0.0352 \\ \hline
    7 & 84 & 0.11 & 0.043 & 0.08 & 0.152 \\ \hline
    8 & 126 & 0.16 & 0.0586 & 0.54 & 0.625 \\ \hline
    9 & 208 & 0.27 & 0.0742 & 3.64 & 2.50 \\ \hline
    10 & 324 & 0.42 & 0.0742 & 23.2 & 10.0 \\ \hline
    11 & 520 & 0.66 & 0.0898 & 151 & 40.0 \\ \hline
    12 & 792 & 1.03 & 0.105 & 933 & 160 \\ \hline
    13 & 1224 & 1.52 & 0.141 & 5900 & 640 \\ \hline
    14 & 1872 & 2.41 & 0.125 & MEM-OUT & MEM-OUT \\ \hline
    15 & 2828 & 3.62 & 0.129 & MEM-OUT & MEM-OUT \\ \hline
    16 & 4290 & 5.55 & 0.145 & MEM-OUT & MEM-OUT \\ \hline
    17 & 6464 & 8.29 & 0.152 & MEM-OUT & MEM-OUT \\ \hline
    18 & 9690 & 12.7 & 0.246 & MEM-OUT & MEM-OUT \\ \hline
    19 & 14508 & 18.8 & 0.199 & MEM-OUT & MEM-OUT \\ \hline
    20 & 21622 & 28.9 & 0.203 & MEM-OUT & MEM-OUT \\ \hline
  \end{tabular}
\end{center}
\vspace{-6mm}
\end{table}

\section{CONCLUSIONS AND FUTURE WORK}
\label{sec:conclusions}

We have described a new graph-based simulation technique that enables
efficient density matrix simulation of quantum circuits. We
implemented this technique in the QuIDDPro/D simulator. QuIDDPro/D
uses the QuIDD data structure to compress redundancy in the gate
operators and the density matrix. As a result, the time and memory
complexity of QuIDDPro/D varies depending on the structure of the
circuit. However, we demonstrated that QuIDDPro/D exhibited superior
performance on a set of benchmarks which incorporate qubit errors,
mixed states, error correction, quantum communication, and quantum
search. This result indicates that there is a great deal of structure
in {\em practical} quantum circuits that graph-based algorithms like
those implemented in QuIDD\-Pro/D exploit.

We are currently seeking to further improve quantum circuit
simulation. For example, algorithmic improvements directed at specific
gates could enhance an existing simulator's performance. With regard
to QuIDDPro/D in particular, we are also exploring the possibility of
using ``read-k'' ADDs and edge-valued diagrams (EVDDs) in an attempt
to elicit more compression. Lastly, we are studying
technology-specific circuits for quantum-information
processing. Optionally incorporating technology-specific details may
lead to simulation results that are more meaningful to physicists
building real devices, particularly with regard to error modeling.

\vspace{-2mm}

\acknowledgments

This work is funded by the DARPA QuIST program, an NSF grant and a DOE
HPCS graduate fellowship.  The views and conclusions contained herein
are those of the authors and should not be interpreted as necessarily
representing official policies or endorsements of employers and
funding agencies.

\end{document}